\documentstyle[twoside,fleqn,espcrc2,epsfig]{article}
\newcommand{\ec}{\end{center}}

\newcommand{\AmS}{\protect\the\textfont2  A\kern-.1667em\lower.5ex\hbox{M}\kern-.125emS}
\title{Searching for a light Higgs in $\Upsilon$ leptonic decays
\thanks{Research under grant FPA2002-00612.
See arXiv:hep-ph/0206156 for a longer version.}}
\author{Miguel Angel Sanchis-Lozano\thanks{Email:Miguel.Angel.Sanchis@uv.es}
\vspace{0.4cm}\\
Instituto de F\'{\i}sica
Corpuscular (IFIC) and Departamento de F\'{\i}sica Te\'orica,
Centro Mixto Universidad de Valencia-CSIC \\
Dr. Moliner 50, E-46100 Burjassot, Valencia (Spain)}

\begin{document} 
\begin{abstract}
        Leptonic decays of vector-states of bottomonium are analized
        searching for a light pseudoscalar 
        Higgs-like
        neutral boson manifesting via an apparent breaking of lepton 
        universality.
\end{abstract}
\maketitle

\section{INTRODUCTION}
Although there are well established lower mass 
bounds for the standard Higgs 
(e.g. from LEP searches \cite{lep}), the situation
may be different in several scenarios and models beyond the 
standard model (SM)  
where such constraints would not
apply, leaving still room for
light Higgs bosons (see \cite{opal,haber01} for example).
Needless to say, 
any possible experimental signal or discovery strategy
of Higgs-like particles should be examined with
great attention.
In this regard, let us remind that
the search for axions or light Higgs in the decays of 
heavy resonances has several attractive features. 
Firstly, the couplings of the former to fermions are 
proportional to their masses and therefore enhanced with respect to 
lighter mesons. Second, theoretical predictions are more reliable, 
especially with the recent development of effective theories like
non relativistic quantum 
chromodynamics (NRQCD) \cite{bodwin}, appropriate 
to deal with such bound states from first principles.

Indeed, intensive searches for a light Higgs-like boson (to be generically
denoted by $\phi^0$ in this paper)  have been
performed according to the so-called Wilczek mechanism \cite{wilczek}
in the radiative decay of vector heavy quarkonia like
the Upsilon resonance (i.e. $\Upsilon{\rightarrow}\ \gamma \phi^0$). So far,
none of all these searches has been successful, but have 
provided valuable constraints on the mass values of light Higgs 
bosons \cite{gunion}.

Nevertheless, in this work I will focus on a possible signal
of New Physics based on the $\lq\lq$apparent'' breaking of  
lepton universality in 
bottomonium decays: {\em stricto sensu}, lepton universality
implies that the electroweak couplings 
to gauge fields of all charged 
lepton species should be the same. According to the interpretation
given in this work, the possible dependence
on the leptonic mass of the leptonic branching fractions 
of $\Upsilon$ resonances below the $B\bar{B}$ threshold (if 
experimentally confirmed by forthcoming measurements) might be viewed
as a hint of the existence of a Higgs of mass about 10 GeV.

\subsection{Searching for a light Higgs in $\Upsilon$ leptonic decays}

Let us write the well known Van Royen-Weisskopf formula including
color, expressing
the leptonic decay width of the 
$\Upsilon(1S)$ vector resonance without neglecting leptonic masses:
\begin{equation}
{\Gamma}_{{\ell}^+{\ell}^-}\ =\ 
4\alpha^2Q_b^2\ \frac{|R_n(0)|^2}{M_{\Upsilon}^2}\ {\times}\ 
K(x)
\end{equation}
where $\alpha\ {\simeq}\ 1/137$ is the electromagnetic fine 
structure constant;
$M_{\Upsilon}$ denotes the mass of the $\Upsilon$ particle and
$Q_b$ is the charge of the bottom quark ($1/3$ in units of $e$);
$R_n(0)$ is the non-relativistic radial wave function of the
$b\overline{b}$ bound state at the origin; finally
$K(x)=(1+2x)(1-4x)^{1/2}$ with $x=m_{\ell}^2/M_{\Upsilon}^2$.
Let us note that $K(x)$ is a decreasing function of $x$:
the higher leptonic mass the smaller decay rate. However, 
such $x$-dependence is quite weak for bottomonium.

In this paper, we will consider the sequential decay
\begin{equation}
\Upsilon\ {\rightarrow}\ \gamma\ \phi^0\ ({\rightarrow}\ \ell^+\ell^-)\ \ \ ;\ 
\ \ \ell=e,\mu,\tau
\end{equation}
It is a continuum radiative transition which in principle permits the
coupling of the bottom quark-antiquark pair into a particle with $J^{PC}$:
$0^{++}, 0^{-+}, 1^{++}, 2^{++}...$.
In the present investigation, we will confine our attention to the two 
first possibilities: a scalar or a pseudoscalar boson.

\subsection{An intermediate spin-singlet $b\bar{b}$ state?}

In a vector resonance like the $\Upsilon(1S)$, the heavy quark 
pair can be in a ${}^3\!S_1$ color-singlet state in the lowest 
Fock state, but the $Q\bar{Q}$ system could also exist with 
other quantum numbers than $J^P=1^-$ since the soft degrees of freedom
can carry the remaining quantum numbers, although with a
smaller probability. These ideas have been cast into the rigorous
formulation of NRQCD \cite{bodwin} and extensively applied 
to heavy quarkonia production and decay. 
Moreover, one can wonder about the possibility of reaching such
Fock states by emission of soft photons instead of soft gluons. 
Let us note however a crucial difference between photons and gluons:
the latter carry color and hence, there exists a lower cutoff corresponding 
to a minimum amount of energy $\lq\lq$taken away'' in the hadronization stage
(corresponding to a pion mass for instance). However, this is not the case
for photons. Actually, the experimental determination of the
leptonic braching fraction (BF) ${\cal B}_{\ell\ell}$ 
actually includes decays accompanied by 
a large number number of soft photons \cite{pdg}.
 
On the other hand, magnetic dipole (M1)
transitions can connect spin-triplet and spin-singlet
states by emission of soft photons from heavy quark lines
(see Fig.1). The probability for this process can be obtained by
dividing the corresponding width \cite{oliver} by by the total width of
the resonance, $\Gamma_{tot}=52.5$ KeV \cite{pdg}, i.e.
\begin{equation}
{\cal P}_{\Upsilon(1S){\rightarrow}\gamma_s(b\bar{b})[{}^1\!S_0]}\ =\ 
\frac{1}{\Gamma_{tot}}\ \frac{4\alpha Q_b^2}{3m_b^2}\ k^3
\end{equation}
where $k$ denotes the energy of the soft photon $\gamma_s$
varying in the range 
$k=10-50$ MeV. Let us also remark that soft photons 
in this experimental context are those
whose energies do not exceed the experimental resolution and hence
are actually not observed \footnote{Typical widths 
of resonance peaks in $e^+e^-$ machines
are of the order of few tens of MeV for the $\Upsilon$
family below open bottom production. Moreover, typical low energy 
cutoffs for photon detection are of the order of 50 MeV.
Notice also that the use of Eq (3) as an estimate for 
the magnetic dipole transition is justified since the
respective wave-lengths of the radiated photons
are quite larger than the size of quarkonium 
(of order $\simeq$ GeV$^{-1}$).}.

\begin{figure}[htb]
\centerline{\hbox{
 \psfig{file=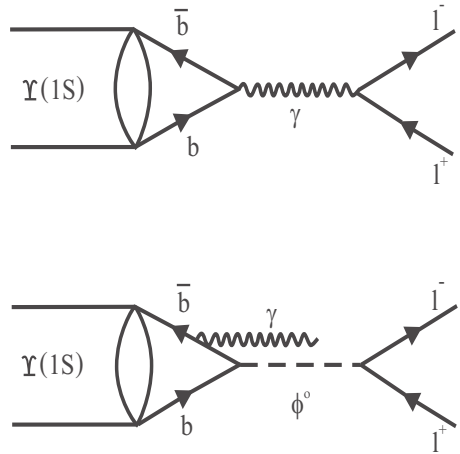,height=6.cm,width=6cm}  
}}
\caption{{\em up:} Electromagnetic annihilation of a $(b\bar{b})[{}^3\!S_1]$ 
bound state (a $\Upsilon(1S)$ resonance
in particular) into a lepton pair via a vector particle (i.e. a photon);
{\em down:} Hypothetical annihilation of a 
$(b\bar{b})[{}^1\!S_0]$ bound state (assuming that the resonance
has previously undergone a 
magnetic transition caused by soft photon emission) 
into  a lepton pair via a Higgs particle.}
\end{figure}

\subsection{Effects of a light $\lq\lq$Higgs''
on ${\cal B}_{\ell\ell}$} 
Let us focus on 
the bottomonium familly of $\Upsilon(nS)$ states below
open flavor (i.e. $n<4$; the $\Upsilon(3S)$ state is however
discarded in the present analysis
since only experimental 
data for the muonic channel \cite{pdg} are currently
available) decaying into a lepton pair plus a soft photon $\gamma_s$: 
\[
\Upsilon(nS){\rightarrow}\ \gamma_s\ b\bar{b}[{}^1\!S_0^{(1)}](\rightarrow 
\phi^0{\rightarrow}\ \ell^+\ell^-)\ \ ;\ 
\ \ell=e,\mu,\tau
\]
where the soft (I stress it once more: {\em unobserved}) 
$\gamma_s$ comes from a  
M1 transition of the $\Upsilon$ resonance,
as sketched in Fig. 1.b. I will
write the decay width ${\Gamma}_{\gamma_s\ell\ell}$
corresponding to the formation of an intermediate state
followed by its annihilation via an scalar or pseudoscalar in
the factored form:
\begin{equation}
{\Gamma}_{\gamma_s\ell\ell}=
{\cal P}_{\Upsilon(1S){\rightarrow}\gamma_s(b\bar{b})[{}^1\!S_0]} \times
\tilde{\Gamma}_{\ell\ell}
\end{equation}
where $\tilde{\Gamma}_{\ell\ell}$ stands for the annihilation width of
the $b\bar{b}$ pair in a
spin-singlet state into a lepton pair via a Higgs ($\phi^0$) boson 
as depicted in Fig.1.
Furthermore, fermions are assumed to couple to the
$\phi^0$ field
according to a Yukawa interaction term in the effective Lagrangian:
\begin{equation}
{\cal L}_{int}^{\bar{f}f}\ =\ -\xi_f^{\phi}\ \frac{\phi^0}{v}
m_f\bar{f}(i\gamma_5)f 
\end{equation}
where $v=246$ GeV stands for the vacuum expectation value 
of the standard Higgs boson; $\xi_f^{\phi}$ denotes a factor
depending on the type of the Higgs boson, which
could enhance the coupling with a fermion 
(quark or lepton) of type $\lq\lq$$f$''. 
Lastly, note that the
$i\gamma_5$ matrix stands only in the case of 
a pseudoscalar $\phi^0$ field.

Now, let us tentatively assume that the mass of the 
light Higgs sought stands 
close to the $\Upsilon(1)$ resonance but below $B\bar{B}$ production:
$m_{\phi^0}\ {\simeq}\ 2m_b$. As will be argued 
from current experimental data in the
next section, I am supposing specifically that $m_{\phi^0}$ lies
somewhere between the $\Upsilon(1S)$ and $\Upsilon(2S)$ masses,
i.e.
\begin{equation}
m_{\Upsilon(1S)}\ {\leq}\ m_{\phi^0}\ {\leq}\ m_{\Upsilon(2S)}
\end{equation}

Next, let us define the mass difference:
${\delta}m=|m_{\phi^0}-m_{\Upsilon}|$, where $\Upsilon$
denotes either a $1S$ or a $2S$ state. Accepting for simplicity that
the Higgs stands halfway between the mass values of both
resonances, ${\delta}m\ {\simeq}\ 0.25$ GeV
for an order-of-magnitude calculation. Hence the scalar tree-level 
$\phi^0$ propagator can be written approximately as
\begin{equation}
\frac{1}{(m_{\Upsilon}^2-m_{\phi^0}^2)^2}\ \simeq\ 
\frac{1}{16\ m_b^2\ {\delta}m^2}
\end{equation}
where the width of the Higgs boson has been neglected for it should
be very narrow due the smallness of $m_{\phi^0}$ and,
moreover, standing below bottom open production.

Finally, one can compare
the relative rates by means of the following dimensionless ratio
\begin{equation}
{\cal R}=\frac{{\cal B}_{\Upsilon{\rightarrow}\gamma_s\ell\ell}}
{{\cal B}_{\ell\ell}}=
\biggl[\frac{m_b^2k^3\xi_b^2\xi_{\ell}^2}
{8\pi^2{\alpha}\Gamma_{tot}v^4}\biggr]
\times \frac{m_{\ell}^2}{{\delta}m^2}
\end{equation}
where we are assuming that the main contribution to the leptonic
channel comes from the photon exchange graph of Fig. 1.a. 
Let us point out once again that since $\gamma_s$ is undetected, 
the Higgs contribution of figure 1.b would be experimentally ascribed
to the leptonic channel of the $\Upsilon$ resonance.

For the sake of a comparison
with other Higgs searches, I will identify
the $\xi_f$ factor with
the 2HDM (type II) parameter for the universal down-type
fermion coupling to a CP-odd Higgs, i.e. $\xi_b=\xi_{\ell}=\tan{\beta}$,
defined as the ratio of the vacuum expectation values of
two Higgs fields 
\cite{gunion}. Inserting
numerical values, 
\begin{equation}
{\cal R}\ \simeq\ (3.6{\cdot}10^{-9}-4.5{\cdot}10^{-7}) \times  
\tan{}^4\beta \times m_{\ell}^2
\end{equation}
where use was made of the approximation
$m_{\phi^0} \simeq 2m_b\ \simeq 10$ GeV, and 
the range $10-50$ MeV for the soft photon energy $k$; $m_{\ell}$
is expressed in GeV.

\section{HYPOTHESIS TEST ON LEPTON UNIVERSALITY}

From inspection of experimental data presented in
Table 1, one realizes a slight but steady increase of the decay rate
with the lepton mass. In spite of that, current 
error bars ($\sigma_{\ell}$) are still too large 
(especially in the case of the $\Upsilon(2S)$) to
permit a thorough check of the lepton mass dependence as
expressed in Eq.(9). 
Nevertheless, I will apply below a hypothesis test
in order to draw, if possible, a statistically
significant conclusion about lepton 
universality breaking. To this end, I present in Table 2
the differences ${\Delta}_{\ell\ell'}$ divided 
by their respective errors ${\sigma}_{\ell\ell'}$, between 
BF's of distinct channels obtained from Table 1. 
Then applying a {\em one-tailed test} \cite{frodesen}, I define
the {\em region of rejection} above a preassigned {\em critical value}
of the ${\Delta}_{\ell\ell'}/\sigma_{\ell\ell'}$ variable
(i.e. positive values if $m_{\ell'}>m_{\ell}$), assuming a
normal distribution.

\begin{table} [htb]
\caption{Branching fractions ${\cal B}_{\ell\ell}$ (in $\%$) of
$\Upsilon(1S)$ and $\Upsilon(2S)$ leptonic decays (from \cite{pdg}).}
\begin{center}
\begin{tabular}{cccc}
\hline
channel: & $e^+e^-$ & $\mu^+\mu^-$ & $\tau^+\tau^-$  \\
\hline
$\Upsilon(1S)$ & $2.38 \pm 0.11$ & $2.48 \pm 0.06$ & $2.67 \pm 0.16$ \\
\hline
$\Upsilon(2S)$ & $1.18 \pm 0.20$ & $1.31 \pm 0.21$ & $1.7 \pm 1.6$ \\
\hline
\end{tabular}
\end{center}
\end{table}
\begin{table} [htb]
\caption{All six differences ${\Delta}_{\ell\ell'}$ (from Table 1) between
the leptonic branching fractions (in $\%$) corresponding to
$\Upsilon(1S)$ and $\Upsilon(2S)$ resonances separately, i.e.
${\Delta}_{\ell\ell'}={\cal B}_{\ell'\ell'}-{\cal B}_{\ell\ell}$; the
${\sigma}_{\ell\ell'}$  values were obtained from Table 1 by summing
error bars in quadrature. Note that only two 
${\Delta}_{\ell\ell'}/{\sigma}_{\ell\ell'}$ for
each resonance can be considered as truly independent.}
\begin{center}
\begin{tabular}{cccc}
\hline
channels & ${\Delta}_{\ell\ell'}$ & $\sigma_{\ell\ell'}$ & 
${\Delta}_{\ell\ell'}/\sigma_{\ell\ell'}$ \\
\hline
$\Upsilon(1S)_{e{\mu}}$  & $0.1$ & $0.125$ & $+0.8$ \\
\hline
$\Upsilon(1S)_{\mu{\tau}}$  & $0.19$ & $0.17$ & $+1.12$ \\
\hline
$\Upsilon(1S)_{e{\tau}}$  & $0.29$ & $0.19$ & $+1.53$ \\
\hline
$\Upsilon(2S)_{e{\mu}}$  & $0.13$ & $0.29$ & $+0.45$ \\
\hline
$\Upsilon(2S)_{\mu{\tau}}$  & $0.39$ & $1.61$ & $+0.24$ \\
\hline
$\Upsilon(2S)_{e{\tau}}$  & $0.52$ & $1.61$ & $+0.32$ \\
\hline

\end{tabular}
\end{center}
\end{table}

The mean of the four ${\Delta}_{e\ell'}/{\sigma}_{e\ell'}$ values
($\ell'=\mu,\tau$ for both $\Upsilon(1S)$ and
$\Upsilon(1S)$ resonances) turns out to be $0.775$. 
Next, I define the {\em test statistic}: 
$T={\langle}{\Delta}_{e\ell'}/{\sigma}_{e\ell'}{\rangle}{\times}
\sqrt{N}=1.55$, where $N=4$ stands for the number of 
independent points. (Note also that we are dealing with a 
Gaussian of unity variance.)
Choosing the {\em critical value} to be ${\simeq}\ 1.3$,  
the lepton universality hypothesis [playing the role  of the {\em null
hypothesis} in our test, predicting a mean zero (or slightly less) value]
can be {\em rejected} at 
a {\em significance level} of 10$\%$ since $T>1.3$ 
Certainly, this result is
not statistically significant enough to make any serious claim 
about the rejection of the lepton universality hypothesis in
this particular process, but points out the interest to
investigate further the alternative hypothesis stemming from Eq.(9).

In order to explain the observed 
${\cal O}(10)\%$ enhancement from the electronic to the
tauonic channel (see Table 1), one gets
\begin{equation}
16\ {\leq}\ \tan{\beta}\ {\leq}\ 54
\end{equation}
depending on the value of $k$, namely from 50 MeV to 10 MeV in (9). A 
caveat is in order: the above interval is purely indicative as it
only takes into account the probability range on the M1 transition 
estimated according to Eq.(3), and not other sources
of uncertainty. 
It is also worthwhile to remark that the interval (9)  
is compatible with the range needed
to interpret the $g-2$ muon anomaly in terms of a
light CP-odd Higgs ($A^0$) resulting from a   
two-loop calculation \cite{cheung}
\footnote{However, after correcting a sign mistake
in the so-called hadronic light by light contribution
in the $g-2$ calculation the discrepancy with
respect to the SM becomes smaller than initially 
expected \cite{knecht}.}.

\section{SUMMARY}
I have pointed out in this paper
a possible breaking of lepton universality 
in $\Upsilon$ leptonic decays, interpreted in terms of 
a neutral CP-odd Higgs of mass around $10$ GeV, introducing a $m_{\ell}^2$
dependent contribution in the partial width. 

I end by emphasizing the interest in more accurate data on
leptonic BF's of $\Upsilon$ resonances, 
particularly considering the exciting possibility of a signal
of New Physics as pointed out in this work.
Hopefully, B factories working below open bottom production will
provide in a near future new and likely more precise measurements of the 
leptonic BF's for the $\Upsilon$ family.
\end{document}